\documentclass[aps,prc,showpacs,preprint]{revtex4-1}
\usepackage{graphicx}
\usepackage[section]{placeins}
\begin{document}
\title{QUANTIZATION OF A ONE-DIMENSIONAL SYSTEM BY MEANS OF THE QUANTUM HAMILTON-JACOBI EQUATION}
\author{Mario Fusco Girard}
\affiliation{Department of Physics ''E.R. Caianiello'',\\University of Salerno \\and \\Gruppo Collegato INFN di Salerno,\\Via Giovanni Paolo II, 84084 Fisciano (SA), Italy\\
electronic address: mario@sa.infn.it}

\begin{abstract}
The numerical version of the Hamilton-Jacobi quantization method, recently proposed, is applied to the one dimensional quartic oscillator. A suitable quantization condition is formulated and various energy levels and wave functions are computed. The results very well agree with those obtained by means of the Schroedinger equation, and confirm that the Quantum Hamilton Jacobi approach, which is the exact version of the semiclassical WKB scheme, is a self-contained quantization procedure, equivalent and independent from the Schoedinger's one but more general. Indeed, with respect to this latter, the Quantum Hamilton-Jacobi equation can be used to investigate the limit $h \to 0$, where the Schroedinger equation loses its significance, and explains how the fundamental quantities of the classical mechanics, the Hamilton's characteristic function and the classical momentum, are generated from the corresponding quantum ones. 
\end{abstract}

\maketitle

\section{Introduction}
A quantization procedure for one-dimensional conservative systems, based on the Quantum Hamilton-Jacobi Equation (QHJE), has been recently presented [1, 2]. The method is independent from the usual quantization scheme based on the Schroedinger equation and gives the same results for the energy levels and the wave functions; in addition, it allows both to exactly represent the wave functions in a WKB-like form inside of the classically allowed regions and permits to understand how the classical characteristic function and momentum are respectively generated, in the classical limit, by the corresponding quantum quantities.

In [1] the numerical construction of the wave functions by means of the QHJE was presented, while in [2] it was shown how, for some Hamiltonians, the same results can be analytically obtained. In both papers, the energy eigenvalues were supposed obtained from the QHJE by means of the procedure devised by Leacock and Padgett [3,4], who showed how it is possible, in some cases, to find the energy levels for one dimensional motion without explicitly solving the QHJE itself. However, as happens for the Schroedinger Equation (SE), when the analytical procedure is not applicable, it is necessary to do recourse to numerical methods to find the energy levels and the wave functions. The present paper aims to show how the QHJE approach works in this case.

The details of the method were presented in the quoted references. Here we will briefly recall the main points of the numerical procedure.

Let us consider a one dimensional motion of a particle of mass $m$ in a potential $V(x)$ with two turning points, $x_1$ and $x_2$ $(x_1 < x_2)$ at the energy E. These points separate the classically allowed region (c. a. r.) from the forbidden ones (c. f. r.) and are very important in the HJ quantization procedure too.
As well known, the time-independent QHJE is obtained when one searches for solutions of the one-dimensional time-independent Schroedinger equation:

\begin{equation}
-{\hbar ^2\over 2m} {d^2\psi\over dx^2}=\left[E-V(x)\right]\psi\ .
\end{equation}    
of the form:
\begin{equation}
\psi(x,E)=Ae^{{i\over \hbar} W(x,E)}
\end{equation}	
as in the usual WKB method [5,6].

The substitution of Eq. (2) in (1) gives the one-dimensional time-independent QHJE:  
\begin{equation}
{1\over 2m}\left({dW\over dx}\right)^2 - {i\hbar \over 2m}{d^2W\over dx^2}=E - V(x)\ .
\end{equation}  
The solutions $W(x, E)$ of this equation are called the quantum characteristic functions or quantum reduced actions [7] of the particle and have a fundamental role in the following. A special solution $W_S (x, E)$ of Eq. (3) is obviously obtained from the complex logarithm of a solution $\psi(x, E)$ of the SE:
\begin{equation}
W_S (x, E) = \hbar [\rm{Arg} \psi - i \log |\psi|]\ .
\end{equation}    
The real part of $W_S(x, E)$ is a staircase function, increasing of $\pi\hbar$ at each node of the eigenfunction. Actually, as shown in [1, 2], besides $W_S (x, E)$ there is a whole family of different solutions of Eq.(3), depending on a parameter, and whose real part varies smoothly.
For $\hbar = 0$, Eq. (3) reduces to the classical Hamilton-Jacobi equation for the Hamilton's classical characteristic function $W_C(x, E)$ [8]: 
\begin{equation}
{1\over 2m}\left({dW_C\over dx}\right)^2 = E - V(x)\ ,
\end{equation}  
whose solutions are 
\begin{equation}
W_C(x,E) = \int p_c(x,E) dx = \pm\int \sqrt{2m \left({E- V(x)}\right)} dx \ ;
\end{equation} 
We search for purely imaginary solutions of Eq. (3) in the classically forbidden regions, where the wave functions exponentially vanish for $|x|\to \infty$, and for complex solutions in the c. a. r., were the wave functions have an oscillating behaviour. Hereafter, these solutions have to be matched together at the turning points, by imposing the continuity conditions for the wave function and its first derivative. This procedure is analogous to the WKB method, with the difference that we make use of exact solutions of the QHJE, not the approximate semi classical ones.

In the c. f. r. the quantum characteristic function $W(x, E)$ is therefore looked for in the form:
\begin{equation}
W(x, E) = i Y(x, E)
\end{equation}  
while in the c. a. r.  it is a complex function:
\begin{equation}
W(x, E) = X(x, E)  +  i Y (x, E)  \ .
\end{equation}   
By inserting the last equation into Eq. (3) we get two equations for the real and the imaginary parts of W(x, E) in the c. a. r. (apices denote derivatives with respect to $x$, and hereafter the dependence on the energy E of the various quantities not always will be explicitly indicated): 
\begin{equation}
X'^2(x) - Y'^2(x) + \hbar Y''(x)  = 2m\left(E -V(x) \right)
\end{equation}   
\begin{equation}
X'(x) Y'(x) -{1\over 2} \hbar X''(x)  = 0 \ .
\end{equation}   
Equation (10) is immediately integrated:
\begin{equation}
Y(x)=\hbar \log\left[\sqrt | X'(x)|\right] +\rm{const} \ .
\end{equation}   
In the c. f. r. the quantity $Y(x, E)$ satisfies the same Eq. (9) but with $X(x, E) = 0$. The wave functions in the c. f. r., which we denote as I (at the left of $x_1$) and III (at the right of $x_2$) have respectively the forms 
\begin{equation}
\psi_{I,III}(x,E)=A_{I,III}e^{-Y_{I,III}(x,E)/\hbar}
\end{equation}   
where $A_I$ and $A_{III}$ denote suitable constants.
By substituting Eq. (11) into Eq. (9) we get a nonlinear third order equation [5] for the real part of the quantum reduced action X(x, E) in the c. a. r.:
\begin{equation}
{4X'^4(x)-3\hbar ^2 X''^2(x) + 2\hbar^2X'(x)X'''(x)\over 4X'^2(x)} =2m(E-V(x))\ .
\end{equation}       
When a solution of this equation, different from the real part of $W_S (x, E)$ in Eq. (4), is known, the wave function in the c. a. r. can be put in the form [1, 2]:
\begin{equation}
\psi_{II}(x)=\frac{A_{II}}{\sqrt{|X'(x)|}}\rm{Sin}\left[{X(x)\over \hbar}+ \alpha\right]\ .
\end{equation} 		     

This representation is obviously not possible by using the staircase function $X_S (x, E) = Re[W_S (x, E)]$. 
According to the usual choice in the WKB method [5, 6], we will assume $X(x_1) = 0$ in the Eq. (14). Then, in order to have a value of the wave function different from zero in $x_1$, a constant $\alpha$ has to be added to $X(x)/\hbar$ in the argument of the sine function; following a correspondence principle, we choose it as $\pi/4$ so that in the semi classical limit the representation (14) reduces to the WKB one, where this choice is dictated by the behavior of the Airy functions near the turning points [6].  Indeed, when $h\to 0$, the imaginary part $Y(x)$ of the quantum reduced action in the c.a.r. goes to zero, according to Eq. (11), and the Eq. (13) becomes the classical HJ Equation for $X(x, E)$: therefore, this latter function becomes the classical reduced action $W_C (x,E)$, as discussed in [2]; its derivative  $X'(x, E)$ tends toward the classical momentum $p_c (x, E)$, and the expression (14) becomes the well known formula for the WKB wave function in the classically allowed region. As for the quantum reduced action in the classically forbidden regions, it remains purely imaginary for $\hbar \to 0$, generating in this way the corresponding imaginary classical quantity.
The representation (14) holds for any one-dimensional potential; moreover it is exact, and differently from the approximate WKB analogous formula, is valid at the turning points too, where instead the WKB expression diverges.
Eq. (14) shows that the real part $X(x, E)$ of the quantum reduced action $W(x, E)$ is a fundamental quantity in quantum mechanics, being the phase of the wave function in the c. a. r, while the derivative $X'(x, E)$ controls its amplitude.

In [1] various wave functions for some typical Hamiltonian, obtained by numerically solving the QHJE, were presented, while in [2] a method to construct the wave functions by means of analytical solutions of the same equation was demonstrated. In both cases, the preliminary determination of energy eigenvalues was assumed; indeed, Leacock and Padgett [3, 4] obtained the following QHJE quantization condition for the energies:
\begin{equation}
\oint p_s(x,E) dx = 2n\pi\hbar\ ,
\end{equation} 
where $p_s(x, E)$ is the derivative with respect to $x$ of the function $W_S (x, E)$ defined in Eq. (4), and the integration is done along a path in the complex $x-$plane, enclosing the turning points. For various systems, the integral (15) can be analytically linked to the energy, and the exact energy levels can so be obtained, without solving the QHJE. For these systems the analytical QHJ method allows often to get the wave functions [2].

In the present paper we will instead suppose that for the system under consideration, this analytical scheme cannot be applied, and we will show how the QHJE method in this case allows to compute both the energy eigenvalues and the wave functions.

The procedure is based on the following quantization condition: a value of E is an energy eigenvalue if with this choice of the parameter it is possible to construct a normalizable wave function, continuous together with its first derivative, by matching at the turning points the functions (12) and (14), separately found in the various regions by integrating the QHJE.  This condition is analogous to the corresponding quantization condition in the Schroedinger approach.
The numerical procedure is very simple: the starting point is to choose a tentative value for the energy E. Then Eq. (9) with $X’(x, E)$ put equal to zero is numerically integrated for the functions $Y_{I,III} (x, E)$, in both the classically forbidden regions I and III, i. e.  between $- \infty$ and $x_1$, and between $x_2$ and $+ \infty$, looking for solutions with the suitable behaviour for $x\to -\infty$ and $x \to +\infty$, respectively. From these, the two functions $\psi_I(x,E)$ and $\psi_{III}(x,E)$, according to Eq. (12) are obtained. The constant $A_I$ in this step can be chosen equal to 1. Thereafter, the equation (13) for $X(x, E)$ is numerically integrated between the two turning points, with the following conditions in $x_1$: $X(x_1, E)$ is put equal to zero,  two of the three parameters $A_{II}, X'(x_1, E)$ and $X''(x_1, E)$ are chosen so that the tentative wave function $\psi_{II} (x,E)$ and its first derivative continuously match in $x_1$ with the previously computed  $\psi_I (x,E)$, and to the third parameter an arbitrary value is given (more on this later). A tentative wave function (14) in the c. a. r. is so built. The next step is at the second turning point $x_2$: the constant $A_{III}$ in eq. (12) is chosen such that the two tentative expressions for the wave functions $\psi_{II}(x,E)$ and $\psi_{III}(x,E)$  in the two regions II and III, respectively, have the same value in $x_2$. After that, if the first order derivatives of the two functions are equal in $x_2$, E is an energy eigenvalue and the wave function represented by:

\begin{eqnarray}
&&\psi_{I}(x,E) = A_{I}e^{-Y_I(x,E)/\hbar}\qquad  \rm{for} \ -\infty <x\leq x_1 \nonumber \\
&&\psi_{II}(x)=\frac{A_{II}}{\sqrt{|X'(x)|}}{\rm Sin}\left[{X(x)\over \hbar}+ {\pi \over 4}\right]\qquad  {\rm for} \ x_1 \leq x \leq x_2  \\
&&\psi_{III}(x,E)= A_{III}e^{-Y_ {III}(x,E)/\hbar}\qquad  \rm{for} \ x_2 \leq x < + \infty \nonumber  
\end{eqnarray} 
is the normalizable eigenfunction corresponding to the energy E, continuous with its first derivative along the whole $x-$axis and the procedure terminates, apart for the fact that the three functions can be multiplied by a normalization factor. In force of the unicity theorems for differential equations, the wave function so constructed is the same as would be obtained from the numerical integration of the SE. Otherwise, if the continuity condition on the first derivative in $x_2$ is not satisfied, the procedure has to be repeated with a different value of the tentative energy E, until the eigenvalue is obtained with the wanted approximation.

The method has been applied successfully to many Hamiltonians, and in the following we report some results for a quartic oscillator, with potential:
\begin{equation}
V(x) = \frac{1}{2}  k x^2 + \lambda x^4
\end{equation}   
The first three eigenvalues, with $k =1$ and for various values of $\lambda$, are presented in the table below (the number $n$ is the number of nodes of the eigenfunction) together with the corresponding values as computed by Hioe and Montroll [9] by means of the SE. As seen the agreement is very satisfying.

\begin{table}[!h]
\centering

\label{my-label}
\begin{tabular}{|l|l|l|l|}
\hline
 &\ $n=0$  &\ $n=1$  &\ $n=2$   \\ \hline
$\lambda = 0.002$ & 0.5014895 (0.50148966) & 1.5074192 (1.50741940) & 2.51920 (2.51920212) \\ \hline
$\lambda = 0.01$  & 0.50725615 (0.50725620) & 1.5356482 (1.53564828) & 2.590842 (2.59084580) \\ \hline
$\lambda = 0. 1$  & 0.5591463 (0.55914633)  & 1.769450 (1.76950264)  & 3.13862431 (3.13862431)\\ \hline
$\lambda = 1.0$  &	0.80377065(0.80377065)	& 2.737789( 2.73789227) & 5.179295(5.17929169) \\ \hline
\end{tabular}
\caption{The first three eigenvalues computed by means of the present method for various values of $\lambda$; into the brackets are reported the corresponding values as computed by means of the SE [9].}
\end{table}

\begin{figure}[!h]
\includegraphics[scale=1]{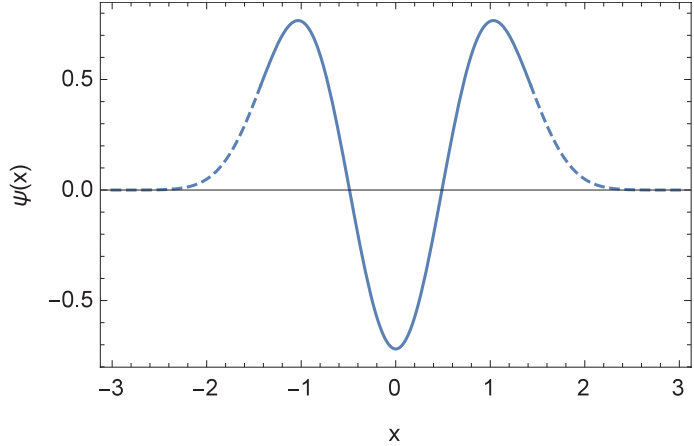}%
\caption{The $n=2$ wave function obtained for the quartic oscillator by the present method, with the value $\lambda = 1.0$ of the coupling constant, is reported; the energy value is 5.179295. The turning points are $x_{1,2}= \pm 1.42811$. The dashed curves represent the functions $\psi_I(x,E)$ and $\psi_{III}(x,E)$ in the two c.f. regions, while the continuous curve is the plot of the function $\psi_{II}(x,E)$, which is the part of the wave function in the c.a.r. As described in the text, the three functions are numerically computed by separately integrating the QHJE in the three regions, and subsequently are matched together at the turning points.}
\end{figure} 

In Fig. 1 the wave function for $\lambda = 1.0$, $n = 2$ and $E = 5.179295$, obtained with the present method for the quartic oscillator, is plotted. The turning points are $x_{1,2}= \pm 1.42811$. The dashed curves represent the functions $\psi_I(x,E)$ and $\psi_{III}(x,E)$ in the two c.f. regions, while the continuous curve is the plot of the function $\psi_{II}(x,E)$, which is the part of the wave function in the c.a.r. As explained above, the three functions are numerically computed by separately integrating the QHJE in the three regions, then are matched together at the turning points. The matching of the functions and the first derivative at the turning points is very good, so demonstrating that the indicated value of the energy E approximates very well the corresponding exact energy eigenvalue for the quartic oscillator. As happens for every wave function so far computed with the method described, this wave function too very well agrees with the corresponding one computed by numerically integrating the SE with the same values of the parameters.

A clarification is in order at this point. As previously said, when integrating Eq. (13) two of the three parameters $A_{II}, X'(x_1, E)$ and $X''(x_1, E)$ are fixed by the continuity conditions of the wave function and its derivative, while the third is arbitrary. It is convenient to choose $b = |X'( x_1, E)|$ as the free parameter. To $b$ every strictly positive value can be given. All the functions $X(x, E, b)$ computed in this way are different but equally exactly reproduce the wave function in the c.a.r. when inserted into Eq. (14). This is due to the fact that Eq. (13) is a third order equation.  When $b\to 0$, the corresponding $X(x, E, b)$ tends to the real, staircase part of the function $W_S (x, E)$ in Eq. (4), whose variation $\Delta X$ between $x_1$ and $x_2$ is $n$ $\pi$ $\hbar$, $n$ being the number of nodes of the wave function. While increasing $b$, this variation increases too. For a particular value $b^*= b^*(E)$, the variation $\Delta X$ is:
\begin{equation}
\Delta X = X(x_2, E, b^*) - X(x_1, E, b^*) = (n+ \frac{1}{2}) \pi \hbar
\end{equation}		
and we consider the corresponding reduced action $W(x, E, b^*)$ as the only physically acceptable solution of the QHJE, different from $W_S (x, E)$. As discussed in [2], with this condition, the phase of the wave function in (14) varies of $\pi$ for each increment of $x$ corresponding to a period, and the wave function changes only by a factor $-1$. The same happens for the WKB wave function, which satisfies the well known quantization condition [5, 6]
\begin{equation}
\Delta W_C = \int ^{x_2}_{x_1} p_c(x,E) dx = \left(n + \frac{1}{2}\right) \pi\hbar 
\end{equation}            
The condition above is obtained from the one in Eq. (18) in the semiclassical limit.

For an even potential like the quartic one in Eq. (17), condition (18) moreover ensures that the derivative $X' (x, E, b^*)$ is an even function of $x$, as the classical momentum $p_c(x, E)$, which is its limit for $\hbar\to 0$.

As discussed in [2], the condition (18) does not disagree with the one by Leacock and Padgett in Eq. (15), due to the fact that they refer to two different solutions of the QHJE at the same energy.

\begin{figure}[!h]
\includegraphics[scale=1]{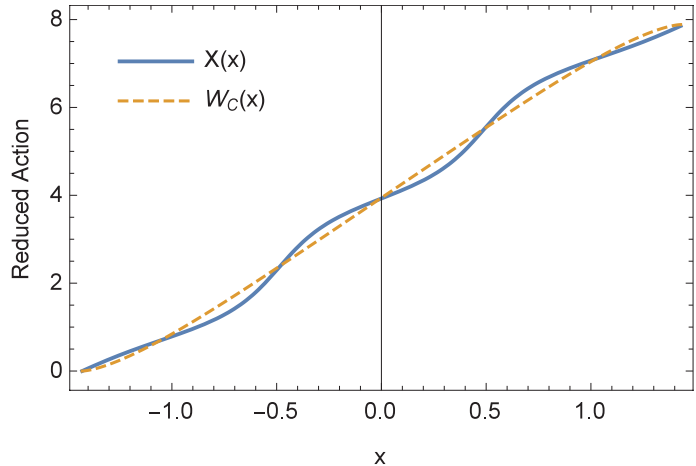}%
\caption{The real part $X(x, E, b^*)$ of the physically acceptable quantum reduced action (thick line) for the wave function in Fig. 1 is reported, together with the corresponding classical quantity $W_C(x, E)$ (thin line). Both these functions are defined only inside the c. a. region. As seen from the figure, the quantum function follows waving the profile of the classical function. The nodes of the wave function are the values of $x$ for which $X(x, E)$ equals $(k-1/4) \pi\hbar$  where $k$ is an integer number.}
\end{figure} 

In Fig. 2 the real part $X(x, E, b^*)$ of the physically acceptable quantum reduced action (continuous line) for the wave function in Fig. 1 between the turning points, is plotted together with the corresponding classical quantity $W_C(x, E)$ (dashed line). In the limit $h\to 0, X(x, E, b^*)$ produces $W_C(x, E)$. As seen from the figure, the quantum function follows waving the profile of the classical function. As shown in [2], this behavior is typical. The nodes of the wave function are the values of $x$ for which $X(x, E)$ equals $(k-1/4) \pi \hbar$  where $k$ is an integer number.

\begin{figure}[!h]
\includegraphics[scale=1]{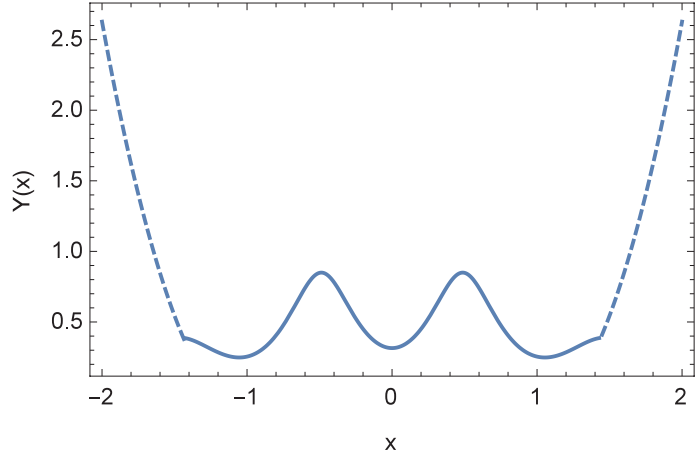}%
\caption{The imaginary parts $Y (x, E)$ of the quantum reduced action for the state in Fig. 1 are plotted. The dashed curves refer to the classically forbidden regions, while the continuous one is the function as computed in the c. a. r. II.}
\end{figure} 

In Fig. 3 are plotted, for the same state, the imaginary parts $Y (x, E, b^*)$ of the quantum reduced action for the three regions I, II and III. The dashed curves refer to the classically forbidden regions, while the continuous part is the function as computed in the c. a. r. II.

\begin{figure}[!h]
\includegraphics[scale=1]{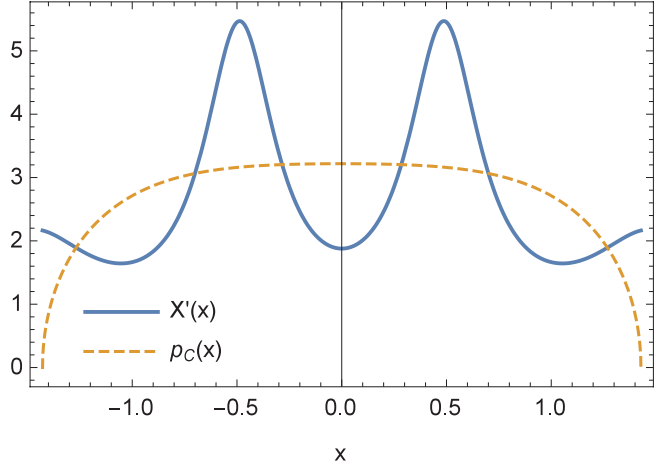}%
\caption{The derivative $X'(x, E, b^*)$ (continuous line) and the classical momentum $p_c(x, E)$ (dashed line) at the same energy as in the other figures are plotted. In the classical limit, the derivative $X'+i Y'$ of the quantum reduced action becomes the classical momentum, in the way discussed in [2].}
\end{figure} 

In Fig. 4 the derivative $X'(x, E, b^*)$ (continuous line) and the classical momentum $p_c( x, E)$ at the same energy and with the same values of the parameters of the previous Figs. are plotted. In the classical limit, the derivative $X' + i Y'$ of the quantum reduced action becomes the classical momentum, in the way discussed in [2].

\begin{figure}[!h]
\includegraphics[scale=1]{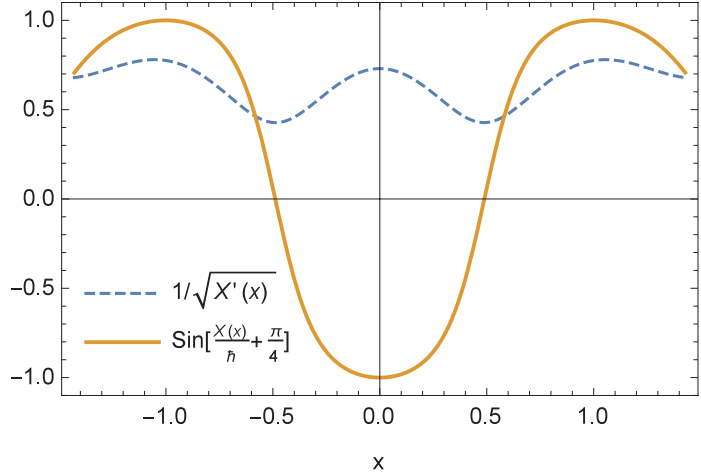}%
\caption{The continuous line is the plot of the function $A_{II} \rm{Sin}\left[{X(x)/ \hbar}+ \pi/4\right]$, while the dashed line is the plot of $1/\sqrt{|X'(x, E)| }$, always for the same state in Fig. 1. The product of these two functions gives the wave function in the c. a. r.}
\end{figure} 

Finally, in Fig. 5, the continuous line is the plot of the function 
\begin{equation}
A_{II} {\rm Sin} \left[{X(x)\over\hbar}+{\pi\over 4}\right]                                                                
\end{equation}   
while the dashed line is the plot of $1/\sqrt{|X'(x, E)|}$, always for the same state. The product of these two functions gives the wave function in the c. a. r., plotted in Fig. 1.

In conclusion, the results presented in this paper, together with those in Refs. [1-4], demonstrate that the QHJE method is a self contained quantization procedure, independent from the SE one. Indeed, in this approach, the QHJE can be postulated and the energy levels and wave functions are analytically [2,3,4,10] or numerically obtained as shown in the present paper. 
With respect to the usual SE approach, the one based on the QHJE gives the same results but is more general, in various respects: firstly, by putting in it $h=0$, the formulation of the classical mechanics based on the Hamilton-Jacobi equation is recovered.  Moreover, the QHJE allows to investigate the limit $h \to 0$, where instead the SE loses its significance. In this limit, the QHJE approach becomes the WKB semi classical quantization method, so that it can be considered as the exact version of this latter.

Finally, this approach illuminates the fundamental role that the quantum reduced action has in quantum mechanics, and how the basic quantities of the classical mechanics, the Hamilton's characteristic function and the classical momentum, are generated from the corresponding quantum quantities.

\end{document}